\documentclass[structabstract]{aa}  
\usepackage{natbib}
\bibpunct{(}{)}{;}{a}{}{,} % to follow the A&A style
\usepackage{graphicx}
\usepackage{txfonts}
\begin{document}
   \title{Massive young disks around Herbig Ae stars\thanks{Based on observations carried out with the
 IRAM Plateau de Bure Interferometer. IRAM is supported by
 INSU/CNRS (France), MPG (Germany) and IGN (Spain).} }

   \author{J. Boissier
          \inst{1,2,3}
	  \and
           T. Alonso-Albi
          \inst{4}
          \and
           A. Fuente
          \inst{4}
          \and
          O. Bern\'e
          \inst{5}
          \and
          R. Bachiller
          \inst{4}
          \and
          R. Neri
          \inst{3}
	  \and
          D. Ginard
          \inst{4}
}
   \institute{
Istituto di Radioastronomia - INAF, Via Gobetti 101, Bologna, Italy (e-mail: boissier@ira.inaf.it)
   \and
   ESO, Karl Schwarzschild Str. 2, 85748 Garching bei Muenchen, Germany
   \and
   Institut de radioastronomie millim\'etrique, 300 rue de la piscine, Domaine universitaire, 38406 Saint Martin d'H\`eres, France.
          \and 
           Observatorio Astron\'omico Nacional (OAN), Apdo. 112,
             28803 Alcal\'a de Henares, Madrid, Spain       
          \and
        Leiden Observatory, Leiden University, PO Box 9513, 2300 RA Leiden, The Netherlands
}
 \abstract 
% 5 {} token are mandatory
%  \abstract
   {Herbig Ae stars (HAe) are the precursors of Vega-type systems and, therefore, crucial objects in planet formation studies. Thus far, only a few
disks associated with HAe stars have been studied using millimetre interferometers. }
   {Our aim is to determine the dust evolution and the lifetime of the disks associated with Herbig Ae stars.}
   {We imaged the continuum emission at $\sim$3~mm and $\sim$1.3~mm of the Herbig Ae/Be stars BD+61154, RR~Tau, VY~Mon and LkH$\alpha$~198 using
the Plateau de Bure Interferometer (PdBI). These stars are in the upper end of the stellar mass range of the Herbig Ae stars (M$_*$$>$3~M$_\odot$).
Our measurements were used to complete the Spectral Energy Distribution (SED). The modelling of the SED, in particular the FIR-mm part, allow us to determine the masses and dust properties of these disks.}
   {We detected the disks associated with BD+61154, RR~Tau and VY~Mon with disk masses of 0.35~M$_\odot$, 0.05~M$_\odot$ and 0.40~M$_\odot$ respectively. The  disk around LkH$\alpha$~198 was not detected with an upper limit to the disk mass of 0.004~M$_\odot$. We detected, however, the disks associated with the younger stellar objects LkH$\alpha$~198--IR and LkH$\alpha$~198--mm that are located in the vicinity of LkH$\alpha$~198. The fitting of the mm part of the SED reveal that the grains in the mid-plane of the disks around BD+61154, RR~Tau and VY~Mon have sizes of $\sim$1--1000~$\mu$m. Therefore, grains have not grown to centimetre sizes  in these disks yet.}
{ These massive (M$_*$$>$3~M$_\odot$) and young ($\sim$1~Myr) HAe stars are surrounded by 
massive ($\gtrsim$0.04~M$_\odot$) disks with grains of micron-millimetre sizes.
Although grain growth is proceeding in these disks, their evolutionary stage 
is prior to the formation of planetesimals. These disks are less evolved than those detected around T Tauri and Herbig Be stars. }
   \maketitle
%
%________________________________________________________________

\section{Introduction}

HAeBe stars, pre-main sequence intermediate mass objects (M$_*$$\sim$2--8~M$_\odot$), share many characteristics with high-mass stars, such as 
clustering, presence of photo dissociation regions (PDR), etc. HAeBe stars are, however, much closer to us and less embedded than higher mass stars, which makes 
the detection of circumstellar disks around these stars  much more feasible, and key for the understanding of the mechanisms of massive star
formation in general. In addition, determining the lifetimes and disk dispersal mechanisms of circumstellar disks is important for planet formation studies.
Disk lifetimes set an upper limit on the time available for the assembly of planetary systems. Determining the details of the gas and dust dispersal
is necessary to establish the inital conditions for planet formation.

We are carrying out a comprehensive search for circumstellar disks around HAeBe stars using the IRAM Plateau de Bure Interferometer (PdBI).
Our final goal is to investigate the properties of the disks around intermediate mass stars to eventually determine
their occurrence, lifetime, and evolution. We have completed the observations toward 11 HAeBe stars.
The results corresponding to the first 7 stars have already been published by \cite{fue+03,fue+06}, \cite{alo+08}, and \cite{alo+09} 
which will be referred to hereafter as Paper I.
Four  stars are presented in the present paper.

The first 7-star sample was mainly composed of Be stars in the upper end of the intermediate-mass stellar range (M$_*$$>$7~M$_\odot$).
Our results showed that the disks associated with these hot stars have masses 
a factor of 5-10 lower than the masses of the disks associated with Tauri (TT) stars. 
Since all these massive stars are very young, $<$1~Myr, we concluded that disk dispersal 
is significantly more rapid at higher stellar masses ($>$7~M$_\odot$) than in the lower mass stars.
This result is also consistent with recent conclusions by \cite{ber+09} based on PAHs emission from disks. They found that the PAH emission
observed toward hot HBe stars ($>$7~M$_\odot$) is associated with the nebula instead of the disk proving either the non-existence or 
the existence of smaller disks around these stars.
Regarding the dust properties, our results proved the existence of large grain sizes ($\sim$1~cm) in the midplane of the disks associated with
these massive, $>$7~M$_\odot$, stars. If planet formation occurs in these stars, it should occur in timescales of a few 10$^5$~yr.

Several groups have been observing the disks associated with Herbig Ae stars in the last decade \citep{tes+03,pie+03,pie+05,pie+06,ham+06,ham10,ise+07,alo+08,cha+08,ban+11}.
These stars have typical ages of a few Myr and stellar masses between 2 and 3 M$_\odot$. Evidences for grain growth have been found in many of these disks \citep[see e.g.][]{tes+03,ise+07,alo+08,ban+11}.
 Thus far, there is no evidence for a differentiated evolution from that of the disks  around T Tauri stars (TTs). No correlation between the disk mass and neither the stellar mass nor the stellar age have been
found in TTs and Herbig Ae stars \citep[see e.g.][]{nat+07,ric+10tau,ric+10rho}.

 These observational results agree with recent photo-evaporation models \cite{gorhol09}
which predict that the disk lifetimes are similar for disks associated to stars with stellar masses in the range of 0.3$-$7~M$_\odot$, 
but rapidly decline for stars of higher masses (Be stars).
However, observational data are still very scarce, specially in the $\sim$3-7~M$_\odot$ stellar mass range where the change between one regime 
and the other occurs.
We present continuum interferometric observations using the PdBI and the subsequent SED modelling of four sources in this stellar mass range. 
Our goal is to get some insights into the physical characteristics of the circumstellar disks around these stars
that are in the borderline between Herbig Ae and Be stars.
\begin{figure}[!]
\includegraphics{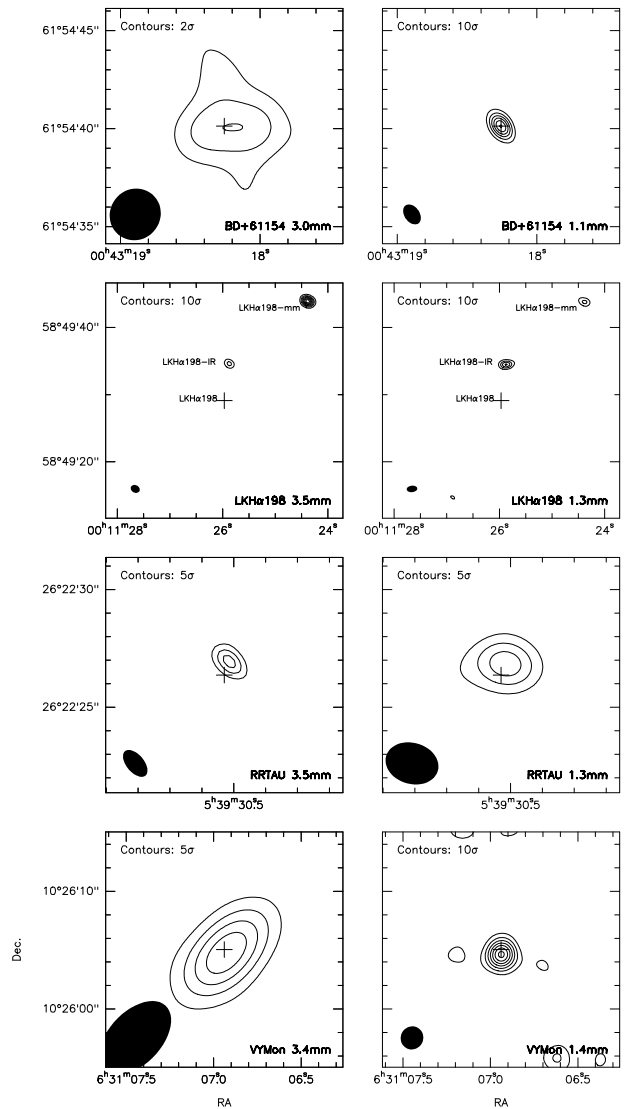}
\vspace{16.0cm}
      \caption{Interferometric maps of the continuum emission of BD+61154, LkH~$\alpha$198, RR~Tau and VY~Mon around $\sim$3~mm (left panels) and $\sim$1.3~mm 
(right panels) as observed with the Plateau de Bure interferometer. 
The exact observing wavelengths are indicated in the lower right corners. The  level spacings (in units of map noise)  are indicated in the upper left corners of the frames and were chosen according the signal to noise ratios of the detections  (see Table~\ref{tab-obs} for the noise values). 
The synthesized  beams are presented in the bottom left corners of the panels.
The beam sizes and observing dates are given in Table~\ref{tab-obs}. 
The measured fluxes  (results of fits in the Fourier plane) are  reported in Table~\ref{tab-fits}.}
         \label{fig-maps}
 \end{figure}

\begin{table*}
\caption{List of sources}             
\label{tab-sou}        

\begin{tabular}{l c c c c c c}
  \hline \hline \noalign{\smallskip}
Star             & Sp.    &  RA(J2000)  & Dec(J2000)  &   d   &   Star Mass  & Age$^1$ \\ 
& & h:m:s          &   $^\circ$:$\arcmin$:$\arcsec$& pc & M$_\odot$ & Myr \\
\hline  \noalign{\smallskip}

BD+61154         & B8     &     00:43:18.257     &   +61:54:40.13"  & 650  & 4.0  & $<$1$^2$    \\
LkH$\alpha$~198     & B9/A5  &  00:11:25.970     &   +58:49:29.10"  & 600  & 3.5  & 1.5  \\
RR Tau           & A2     &     05:39:30.502     &   +26:22:26.90"  & 800  & 3.6  &  1.3  \\
VY Mon           & B8     &     06:31:06.937     &   +10:26:05.04"  & 800  & 6.9  &  0.2 \\  \hline \noalign{\smallskip}
\end{tabular}              

\noindent
$^1$ Stellar ages calculated using the Siess tracks and the stellar data in Table A.13 of \cite{alo+09}.

\noindent
$^2$ See discussion in Sect 5.1.
\end{table*}
\section{Observations}

We carried out interferometric observations in the continuum at $\sim$1.3~mm and $\sim$3.0~mm toward the 4 stars listed in Table~\ref{tab-sou}. 
These objects are HAeBe stars with stellar masses of  $\sim$3--7 M$_\odot$ located at a distance of 600--800~pc
from the Sun. All the sources were observed in 2008-2009 using the PdBI in different configurations.
A summary of the observations is presented in Table~\ref{tab-obs}, including dates, wavelengths and beam sizes.
The data reduction and analysis were performed using the GILDAS software package.
The continuum maps presented in the following were built summing all the channels free of line emission in the data.
Natural weighting has been applied to the measured visibilities.

\begin{table}
\caption{Plateau de Bure observations}             
\label{tab-obs}              
\begin{tabular}{l c c c c}  \hline \hline \noalign{\smallskip}
Source       & $\lambda$  &  Beam size   &   Date        &  rms \\
       &  mm &  $\arcsec$  &        &  mJy/beam\\
 \hline  \noalign{\smallskip} 
BD+61154   & 1.1            &  1.12$\times$0.7        &   Dec. 2008   &  0.5     \\
           & 3.0            &  2.68$\times$2.54       &   Apr. 2008   &  0.3   \\
LkH$\alpha$ 198 & 1.3       &  1.56$\times$0.97        &   Mar. 2009 &  0.5      \\
           & 3.5            &  1.40$\times$1.11        &   Jan. 2009  &  0.1      \\
RR Tau     & 1.3            &  2.27$\times$1.75        &   Apr. 2008  &  0.8    \\
           & 3.5            &  1.35$\times$0.74        &   Feb. 2009  &  0.1    \\
VY Mon     & 1.4            &  1.98$\times$1.89        &   Oct. 2009  &  0.6     \\
           & 3.4            &  7.50$\times$4.45        &   Aug. 2009  &  0.4    \\\hline \noalign{\smallskip}        
\end{tabular}
\end{table}

\section{Results}
The observed $\sim$1.3~mm and $\sim$3.0~mm continuum images are shown in Fig.~\ref{fig-maps}. We detected 3 out of the 4 targets observed with the PdBI. We measured the fluxes by fitting different source models to the visibilities in the Fourier plane ($uv$-plane). 
We selected the source models according to the signal to noise ratio (S/N) of the observations : point source for lower S/N, Gaussian distribution (circular or elliptical depending on the source) for higher S/N values. 
All the fit results (positions, fluxes and width for Gaussian fits) are reported in Table~\ref{tab-fits}.

BD+61154 was previously observed with OVRO at 2.7~mm by Mannings \& Sargent who measured a flux of 11.2~mJy \citep[these observations were quoted in the review by][]{nat+00}.
We measured a flux of 1.7$\pm$0.3~mJy at 3.0~mm using the PdBI. 
This large difference in the flux is probably due to the different frequencies (a factor of 1.4 assuming a spectral index of 3) and 
  to the higher angular resolution of our observations that filter out the extended emission. 
\cite{wen95} published a 1.3~mm
continuum flux of 60~mJy based on single-dish observations carried out with the 30m telescope in this same source. Our interferometric observations recover $\sim$50\%
of the continuum single-dish flux. These young stars are still embedded in the parent core and the single-dish emission is frequently dominated by the emission from the envelope (see Paper I).
According to our fit of the $\sim$1.3~mm data, the size of the emitting region is  $\sim$260~AU.

We did not detect continuum emission toward LkH$\alpha$~198 with an rms of 0.1~mJy/beam at 3.5~mm and 0.5~mJy/beam at 1.3~mm. 
Our upper limit to the $\sim$3~mm continuum flux is a factor of 20 lower than that obtained by \cite{dif+97}. Within the field of view of the LkH$\alpha$~198 image, we detected two millimetre peaks that correspond to the mid-IR source LkH$\alpha$~198--IR 
previously detected by \cite{lag+93} and the submillimetre protostar LkH$\alpha$~198-mm detected by \cite{sanwei94}. LkH$\alpha$198-mm is located at the border of the primary beam at 1.3~mm and for this reason, the flux is quite uncertain.

Our interferometric measurements toward RR Tau are the only ones at 
millimetre wavelengths. 
 We detected a flux of 17.9~mJy at 1.3~mm and flux of  2.1~mJy at 3.5~mm and estimated the size of the emission region to $\sim$720~AU from the 1.3~mm data.

\cite{hen+98} measured a flux of 112 mJy for the dense core in VY Mon using the IRAM 30~m telescope. Our 1.3~mm PdBI
observation gives a flux of 60.4$\pm$1.4~mJy, therefore our interferometric observations filtered out about 50\% of the core flux.  Our 1.3~mm data suggest a size of   $\sim$980~AU for the emitting region in this source.

All the sizes derived above ($\sim$260~AU for BD+61154, $\sim$720~AU for RR~Tau, and $\sim$980~AU for VY~Mon)  are typical for circumstellar disks around intermediate mass stars.  Although some emission from the envelope can be detected with our interferometric observations, the
small source sizes confirm that the detected emission arises mainly from the disk.
This interpretation is also reinforced by the fact that the SEDs we have constructed with our interferometric fluxes are well fitted assuming that all the emission detected in interferometry arises from the disk (envelope emission is negligible).

 A fraction of the interferometric flux measured at $\sim$1.3~mm and $\sim$3.0~mm could be due to the stellar wind emission instead of the dust thermal emission from the disk. 
We have estimated the contribution of the free-free emission 
at millimetre wavelengths by extrapolating previous observations at centimeter wavelengths assuming the power-law 
$F_{\nu}$~$\propto$~$\nu$$^{+0.6}$ \citep[e.g.][]{oln75}. The spectral index, $\alpha$~=~$+$0.6, is
expected for an optically thick isotropic wind. This value is fully consistent with those measured in Herbig Be stars using VLA observations by \cite{alo+08,alo+09}.
It is also consistent, within the errors, with the values derived by 
\cite{ski+93}
in most HAeBe stars, specially 
towards the most massive stars of their sample. Only in two
cases, V645 Cyg and TY Cra, the spectral index clearly differs from this value. The emission towards V645 Cyg was over-resolved and the derived spectal index
was $\alpha$~$\sim$~-0.2$\pm$0.3. 
This value is consistent with optically thin free-free emission or the emission from an HII region.
In the case of TY Cra, a negative spectral index of $-$1.2 is found as well as signs of variability pointing to a non-thermal emission mechanism \citep[see also][]{lom+09}.
\cite{nat+04} found $\alpha$~$\sim$~$+$0.6 in HD~163296 and a value of $\sim$0.0 for HD~35187. We consider that 
$\alpha$~$\sim$~$+$0.6 is a reasonable assumption to estimate the possible contribution of the gas continuum emission at millimetre wavelengths.

In Table~\ref{tab-fits}, we show the millimetre spectral index  derived from our data after subtracting the free-free emission, when its contribution is not negligible.The millimetre spectral index is a good indicator of the grain sizes in the disk midplane. The values in Table~\ref{tab-fits}
are larger than those found in the massive stars reported in Paper I,
 suggesting that the grain sizes are lower: we are dealing with a class of less evolved disks.

\begin{table*}
\caption{Results of the fits to the visibilities. Uncertainties on the output parameters are given in parenthesis.}             
\label{tab-fits}              
\begin{tabular}{l l c c c c c c c}  \hline \hline \noalign{\smallskip}
Source       &  Function$^1$      &  $\lambda$  &  \multicolumn{2}{c}{Source Position}   &  HPW   & PA  & Flux   & Spec. Index    \\ 
       &        &  mm &  RA(J2000)   &  Dec(J2000)   &  $\arcsec$  & $^\circ$  & mJy   &     \\ \hline \noalign{\smallskip}
BD+61154   & E-GAUSS        &  1.1            &  00:43:18.257(0.001)   &  61:54:40.13(0.01)   &  0.39(0.06)$\times$0.19(0.06) & 27(13) & 33(1)     & 2.9(0.6) \\
           & POINT          &  3.0            &  00:43:18.160(0.028)   &  61:54:39.90(0.19)   &                               &        & 1.7(0.3)  &  \\
RR~Tau     & C-GAUSS        &  1.3            &  05:39:30.516(0.004)   &  26:22:26.83(0.05)   &   0.9(0.2)                    &        & 17.9(1.7) & 2.2(0.4) \\
           & C-GAUSS        &  3.5            &  05:39:30.514(0.003)   &  26:22:26.94(0.04)   &   0.3(0.1)                    &        & 2.1(0.2)  &    \\
VY Mon     & E-GAUSS        &  1.4            &  06:31:06.938(0.001)   &  10:26:04.61(0.01)   &  1.22(0.06)$\times$0.77(0.07) & -21(6)  & 60.4(1.4) & 4.0(0.2)$^2$\\
           & POINT          &  3.4            &  06:31:06.926(0.003)   &  10:26:04.72(0.04)   &                               &         & 9.7(0.2)  &  \\
\hline   \noalign{\smallskip}     
\end{tabular}

\noindent
$^1$ The fit of  the Fourier Transform of point source in the $uv$-plane has 3 free parameters: the source RA and Dec positions relative to the phase center and its flux. Fitting a  Gaussian distribution has additional parameters: the width (HPW) for circular Gaussian  (C-GAUSS) and  the major and minor axis for an elliptical Gaussian  (E-GAUSS), as well as its orientation with respect to the north (Position angle PA). 

\noindent
$^2$ This value has been derived after subtracting the contribution of the free-free emission (see text, Sect.~\ref{sec-vymon}).
\end{table*}

\begin{figure*}
 \resizebox{\hsize}{!}{\includegraphics[angle=-90]{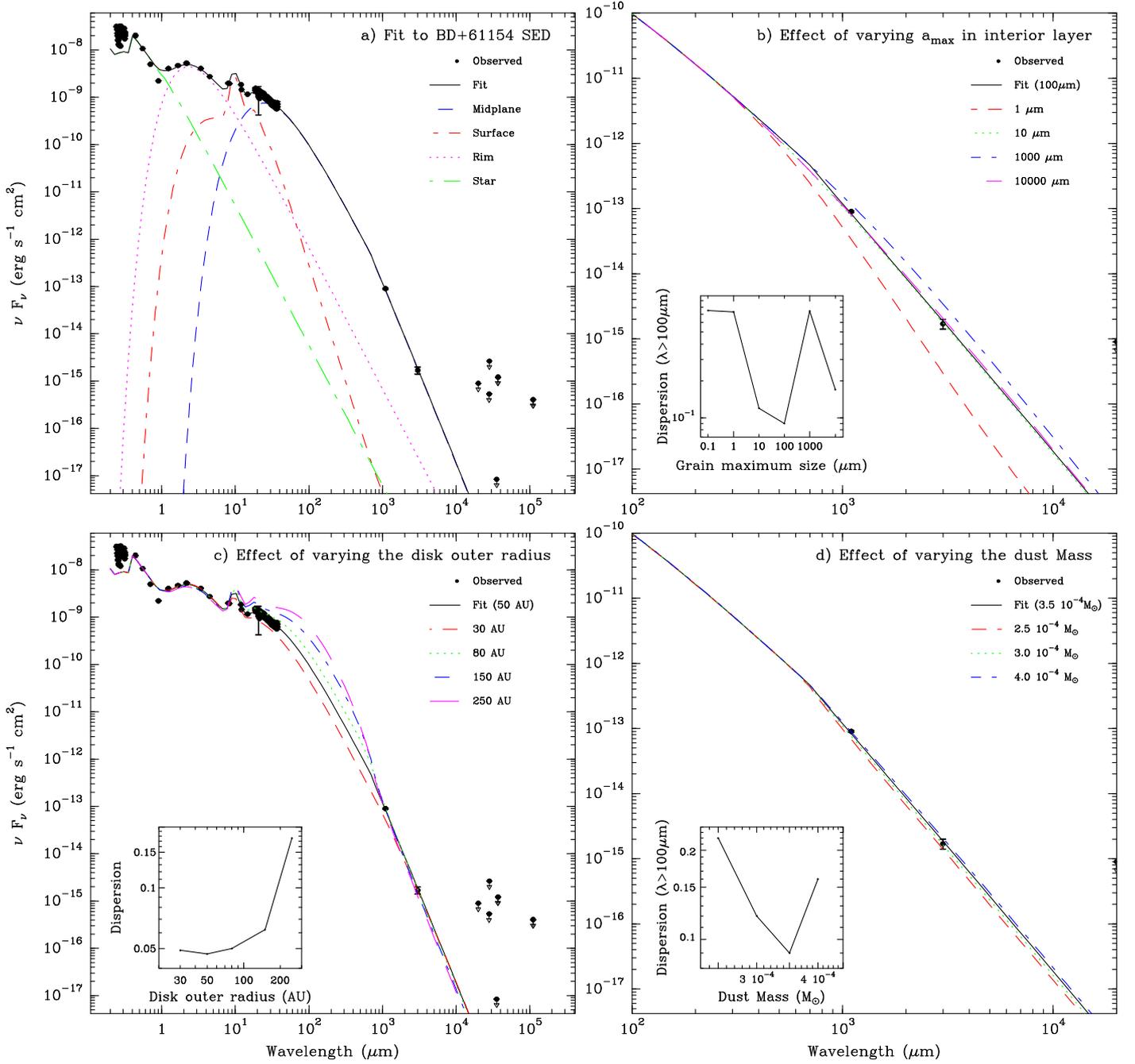}}
      \caption{{\bf a)}~The model fitted to the SED of BD+61154 is shown with all the model contributions: 
interior layer  (blue medium sized dashes), surface layer  (red short and medium sized dashes), inner rim  (pink dotted line), star  (green short and large sized dashes), and total sum of
previous contributions  (black solid line). {\bf b)}~This panel illustrates how the SED changes when one varies the value of the maximum grain size in the interior layer (a$_{max}$).
  The inset show the projected dispersion curve (see Eq.~\ref{eq-disp} for the definition of the dispersion $\chi$)  when only a$_{max}$  varies, all other parameters being fixed at their best fit values.
{\bf c)}~The same as {\bf b)}, but varying the outer radius. 
{\bf d)}~The same as {\bf b)}, but varying the disk mass. The x axis of panels  {\bf b)} and {\bf d)} has been reduced to $\lambda$ $>$ 100 $\mu$m, were the SED is more affected by a change of the parameters a$_{max}$ and disk mass, respectively.
}
         \label{fig-bd}
 \end{figure*}

\section{Modelling the SED}
\label{sec-disp}
We constructed the whole SED toward our sample by combining our PdB data with other results available in the literature and Spitzer photometry fluxes taken in the public archive.
These source fluxes were extracted in a circular aperture of 10$\arcsec$ radius.
Since these stars are bright, we took an off-source flux far away ($\sim1\arcmin$) from the source to avoid contamination. 
The off, containing zodiacal light and ISM contamination is  subtracted to the on-source flux.
The accuracy is about 10\% for this method.
 
At optical and NIR wavelengths, observations are heavily affected by extinction. To correct this reddening we subtract the standard B-V color corresponding to the spectral type of each star, as given by \cite{joh66}, to the observed B-V color. This color excess \textit{E} is used to calculate the visual extinction \textit{A$_v$ = R$_v$ E} assuming a ratio of total to selective absorption $R_v$ = 3.1 \citep{car+89}. With these two parameters we estimate the extinction corresponding to a particular observation using the parameterized extinction law of \cite{car+89}, which can be applied for wavelengths between 0.12 and 3.5 $\mu$m. 
Most HAeBe stars are well known variable stars, a problem that becomes relevant in our case since we are using photometry recovered from the literature in the last decades. The variability affects the B-V color and produces some additional uncertainty in the extinction estimates. 
The disks have been fitted using the passive irradiated circumstellar disk model described by \cite{alo+08} that allows us to consider different grain populations for the midplane and surface layers. 
Grain populations are characterized by the silicate/graphite mixture, the maximum grain size ($a_{max}$), and the slope of the grain size distribution ($n$). 
 A standard grain mixture (SM: 86\% silicate and 14\% graphite) and a value of $n=3.5$ are assumed for the midplane and surface layers.
We allowed $a_{max}$ to vary between 0.1~$\mu$m and 1~cm in the midplane layer.  Observations at millimetre and centimetre wavelengths are not sensitive to grains of sizes larger than a few centimetres.
 During the sedimentation process, large grains ($>$1 mm) are expected to accumulate in the midplane while
the surface layer remains populated by smaller grains. For this reason, in the surface layer we only consider the values $a_{max}$=0.1, 1, 10, 100 $\mu$m.
 The gas to dust ratio is assumed to be 100.
 In addition to $a_{max}$, we vary the inclination angle of the disk ($i$, in the range 10--80$^{\circ}$), the rim temperature ($T_{\rm rim}$, in the range 1200--2000~K, the sublimation temperatures of silicates and graphite being 1500 and 2000~K, respectively) and the index of the density law ($p$, in the range -2.5--0). 
 Finally,  the disk outer radius ($R_{\rm out}$)  and dust mass ($M_{\rm disk}$) vary in  ranges  defined according to a first guess fit of the SED of the sources.
These ranges are, respectively :
30--200~AU and 1.0$ \times 10^{-4}$--10.0$ \times 10^{-4}$~M$_{\sun}$ for BD+6154; 
30--150~AU and 2$ \times 10^{-4}$--32.0$ \times 10^{-4}$~M$_{\sun}$ for RR~Tau;
100--500~AU and  2.0$ \times 10^{-4}$--50.0$ \times 10^{-4}$~M$_{\sun}$ for VY~Mon.

 The goodness of the fit is estimated with a dispersion parameter ($\chi$) defined as: 
\begin{equation}
\label{eq-disp}
\chi~=~\sqrt{\frac{1}{n}\sum{\frac{(F_{observed}-F_{modeled})^2}{{F_{observed}}^2}}}
\end{equation}
where \textit{F} is the flux and where the sum is extended for all detections in the SED from 3~$\mu$m and beyond, being \textit{n} the number of them  (upper limits are not considered). 
Assuming that flux uncertainties are dominated by calibration errors, which we assume to be constant all over the SED  (and of the order of 10\%), this dispersion is proportional to the standard  $\chi$.
The same dispersion definition was used in Paper I.
We report in Table~\ref{tab-mod}   the best fit parameter set, which corresponds to the minimum value of $\chi$, for the three sources we detected with the PdBI.
Table~\ref{tab-mod} contains as well the limits (Min and Max values) of the intervals within which  $\chi$ is lower than $\sim$1.3~ $\chi_{min}$.
 They were calculated by varying each parameter separately while keeping  all the other to their best fit values.  

\begin{figure*}
 \resizebox{\hsize}{!}{\includegraphics[angle=-90]{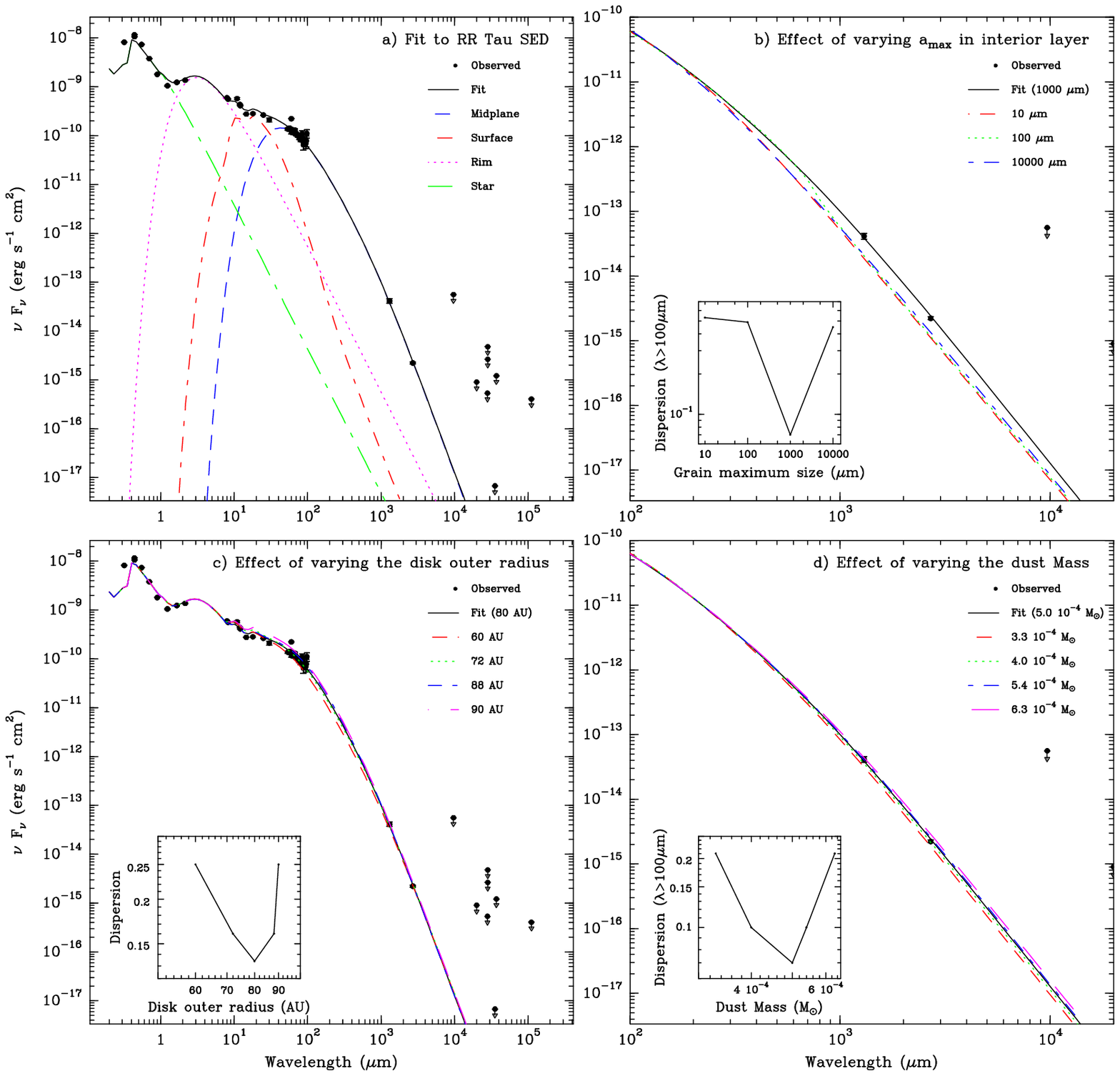}}
      \caption{Same figure as Fig.~\ref{fig-bd} for RR~Tau.}
         \label{fig-rr}
 \end{figure*}

\begin{figure*}
 \resizebox{\hsize}{!}{\includegraphics[angle=-90]{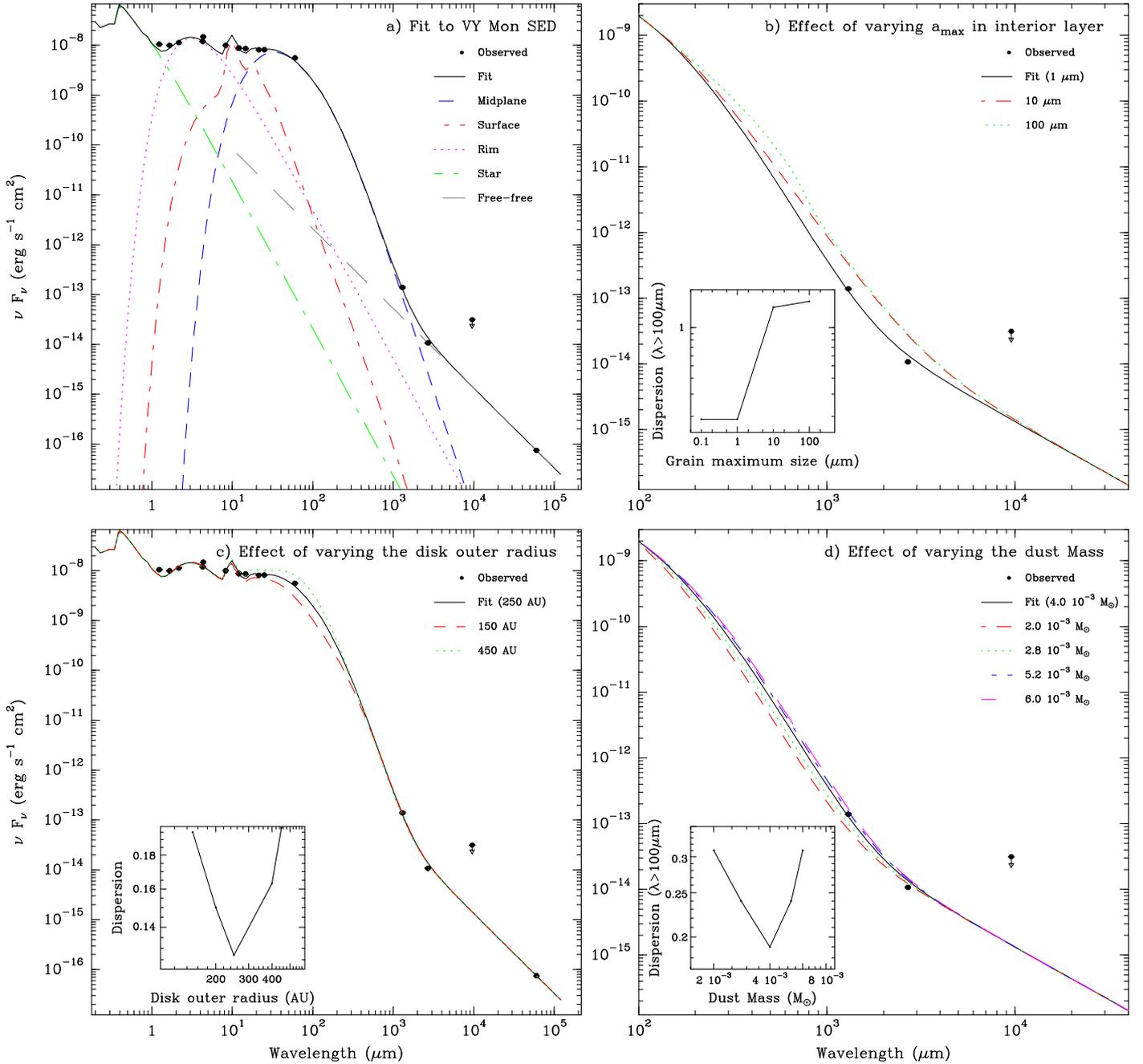}}
      \caption{Same figure as Fig.~\ref{fig-bd} for VY~Mon. Note that in this case we also consider the free-free emission component  (grey large sized dashes).}
         \label{fig-vy}
 \end{figure*}

As M$_{disk}$ and $a_{max}$ affect only the fluxes in the FIR and mm ranges, where we have only few measurements (sometimes only our two PdBI fluxes at $\sim$1.3~mm and $\sim$3.0~mm), their changes impact faintly the normalized dispersion as defined in equation~\ref{eq-disp}.
To enhance their impact, we estimated their Min and Max values  considering only the restricted  SED range  of $\lambda > 100 \mu$m. 
Our fits are shown in Figs.~\ref{fig-bd} to \ref{fig-vy}. In the following we comment the individual sources and the uncertainties in our estimates.

\section{Individual sources}
\label{sec-ind}
\subsection{BD+61154}
BD+61154  (also  designated as MWC 419 or V594 Cas) has spectral
type B8 \citep{her60}, luminosity 330~L$_\odot$ \citep{hil+92}, and is at a distance of 650~pc \citep{hil+92}.
\cite{tes+98} estimated an age of 0.1~Myr assuming T$_{eff}$=11220~K and a luminosity of 300~L$_\odot$.
Later, \cite{alo+09} derived an age of 1.4~Myr adopting a lower value of the luminosity of
220~L$_\odot$. This low value of the luminosity cannot fit the NIR part of the SED. 
We need a luminosity of $\sim$300~L$_\odot$ to fit the SED, which implies that the star is younger than 1~Myr (see Table~\ref{tab-sou}). 

Based on previous VLA observations published by \cite{ski+93}, we estimated that the maximum contributions of the free-free 
emissions are 0.44 and 0.8 mJy at 3.0 and 1.1~mm respectively, which is negligible in comparison with the fluxes measured at the PdBI.
Consequently, the free-free emission is not taken into account in the fit of the SED for BD+61154. 
In Fig.~\ref{fig-bd}a we show our best fit to the SED of this star. The bump appearing in the SED between 1 and 4 microns is well reproduced assuming a rim 
temperature of 1800~K.
The agreement between our model and observations is quite good and allows us to constrain some disk parameters.
The best fit is obtained with an inclination of 70$^\circ$, an outer radius of 50~AU and a density distribution slope of $-$0.7.
Our millimetric data constrain the dust disk mass to $3.5\times 10^{-4}$~M$_{\odot}$ and the dust maximum grain size to 100~$\mu$m in the midplane layer.
Within the errors, the radius of the disk obtained by fitting the SED of BD+61154 is consistent 
with that derived from the fitting of the PdBI $\sim$1.3~mm continuum emission ($\sim$125~AU), which reinforces our interpretation that the interferometric millimetre fluxes originate in the disk.

One has to keep in mind  however that the outer radius is determined by the FIR part of the SED and we have very few
flux measurements in this region (see Fig.~\ref{fig-bd}). Photometry in the 100--1000~$\mu$m wavelength range is necessary to have
a more accurate estimate of the outer radius. 

\begin{table}
\caption{Disk Models}             
\label{tab-mod}              
\begin{tabular}{l l l l l}  \hline \hline \noalign{\smallskip}
BD+61154 & Best fit & Min value & Max Value & Dev$^1$ \\
\hline \noalign{\smallskip}
R$_{out}$ (AU)& 50 & undef$^2$& 150& $\sim$1.3$\chi_{min}$ \\ 
Dust Mass (M$_{\odot}$) & 3.5 10$^{-4}$ & 3.0 10$^{-4}$ & 3.8 10$^{-4}$  & $\sim$1.3$\chi_{min}$ \\
Incl ($^{\circ}$)& 70& 62& 73&  $\sim$1.3$\chi_{min}$ \\ 
p & -0.7& -1.0& -0.3& $\sim$1.6$\chi_{min}$ \\
a$_{max}$ int ($\mu$m) & 100& 10& 1000& $>$1.3$\chi_{min}$ \\
a$_{max}$ surf ($\mu$m) & 0.1& undef$^3$& 100& $>$1.3$\chi_{min}$ \\

\hline
\noalign{\smallskip}
RR Tau & Best fit & Min value & Max Value & Dev$^1$\\
\hline
\noalign{\smallskip}
R$_{out}$ (AU) & 80& 72& 88&  $\sim$1.3$\chi_{min}$\\ 
Dust Mass (M$_{\odot}$) & 5.0 10$^{-4}$ & 4. 10$^{-4}$ & 5.4 10$^{-4}$ & $\sim$1.3$\chi_{min}$ \\
Incl ($^{\circ}$)& 70& 66& 74&$\sim$1.3$\chi_{min}$ \\
p& -1.5& -1.75& -1.25&$\sim$1.3$\chi_{min}$ \\
a$_{max}$ int ($\mu$m) & 1000& 100& 10000& $>$1.3$\chi_{min}$ \\
a$_{max}$ surf ($\mu$m) & 10& 1& 100& $>$1.3$\chi_{min}$ \\

\hline
\noalign{\smallskip}
VY Mon & Best fit & Min value & Max Value  & Dev$^1$ \\

\hline
\noalign{\smallskip}
R$_{out}$ (AU)& 250 & 150 & 450 & $\sim$1.3$\chi_{min}$\\ 
Dust Mass (M$_{\odot}$) & 40 10$^{-4}$ & 28 10$^{-4}$  & 52 10$^{-4}$ &  $\sim$1.3$\chi_{min}$ \\
Incl ($^{\circ}$)& 40& 30 & 50& $\sim$1.3$\chi_{min}$ \\ 
p& -1& -1.5& -0.5& $\sim$1.3$\chi_{min}$ \\
a$_{max}$ int ($\mu$m) & 1& undef$^3$& 10& $>$1.3$\chi_{min}$ \\
a$_{max}$ surf ($\mu$m) & 0.1& undef$^3$& 1& $>$1.3$\chi_{min}$ \\
\hline 
\noalign{\smallskip}
\end{tabular}

\noindent
$^1$ Maximum deviation allowed from the minimum value of $\chi$ to estimate the Min and max values.

\noindent
$^2$ We have not considered values lower than 30~AU for the outer radius of the disk around BD+61154.. 

\noindent
$^3$ We have not considered grains with a$_{max}$~$<$~0.1~$\mu$m. 
\end{table}

\subsection{RR Tau}
RR~Tau is a highly variable HAeBe star also classified as UVOR-type \citep{the+94,oud+01,rod+02},  making  difficult the estimate of  its stellar parameters.
It has been classified between B8-A4 by different authors, with the most recent estimate as A2  \citep{her+04}. Even more uncertain is its distance. 
\cite{tes+98} gave a value of 800~pc. 
\cite{blodji06} modeled the UV spectra and derived a distance of 600~pc. 
More recently, \cite{mon+09} modeled the SED at UV and NIR wavelengths and derived a distance of $\sim$2100~pc.
The authors discussed, however, that the obtained value is very dependent on the assumed extinction and the distance could be lower if a circumstellar disk blocks the stellar light.  In this paper, we have adopted a distance of 800~pc and T$_{eff}$=9000~K that seems a good compromise taking into account the data in the literature. 
The values of disk mass and dust spectral index derived from our modelling do not strongly depend  on on the stellar effective temperature  providing it is between 8000~K and 10000~K as expected for a B6-A4 star.
The distance, however, has an important impact on the disk mass estimate, as it scales as D$^2$.

Using a distance of 800~pc, we got a satisfactory fit of the SED.  
Given the upper limit on the continuum flux at 3.6~cm \citep{ski+93} and the 
spectral index of +0.6 for the free-free emission, it could account for at most
 0.3~mJy at 3.5~mm and 0.6~mJy at 1.3~mm, which would not change significantly the SED fitting process. 
Therefore, we did not include it in the fit. 
The bump appearing in the SED between 1 and 2 microns is well reproduced assuming a rim temperature of 1200~K. 
The geometry of the disk is constrained by the mid infrared part of the SED and the best agreement 
between our model and the observations is obtained for a disk outer radius of 80~AU, an inclination of 70$^{\circ}$ 
and a dust density slope p=$-$1.5.
Our interferometric measurements at millimetre wavelengths are the only ones that are not upper limits above 500~$\mu$m.
They allow us to estimate the dust mass of the disk, M$_{dust}$=$5\times 10^{-4}$~M$_{\odot}$ and the maximum size of dust grains in the midplane, a$_{max}$=1000~$\mu$m. 
The measurements performed by IRAS and ISO around 200~$\mu$m are not used in our fit because given their large beams,
the fluxes measured in these observations are likely dominated by the cloud emission.
Assuming a standard gas/dust ratio of 100, the total dust+gas mass of the disk is 0.05~M$_{\odot}$. 
The size of the disk as determined by fitting the SED is smaller than that derived from the millimetre interferometric observations. The large dispersion of
the flux measurements between 100~$\mu$m and 1000~$\mu$m cast some doubts about the accuracy of this estimate. On the other hand,
the disk could have a complex geometry in the outer part that it is not well described by our simple model. The existence of spiral arms and rings are not unusual in HAeBe stars (AB Aur: \citealt{pie+05}; MWC297: \citealt{alo+09})

\subsection{VY Mon}
\label{sec-vymon}
VY Mon is a B8 star located at a distance of 800~pc. It is the most massive star in our sample. We have not been able to find a good fit for $\lambda$~$<$~1~$\mu$m
probably because the poorly known extinction and photospheric fluxes in this variable star. This, however, does not affect our fit to the disk since
 the stellar effective temperature and the distance are well known.
The free-free emission that was detected at 6~cm by \cite{wen95} corresponds to a flux of 8.4 mJy at 3 mm after  extrapolating with a 0.6 spectral index. 
This flux dominates the emission we observed with the Plateau de Bure.
At 1~mm, the free-free emission contributes to 14.3~mJy, i.e. a third of the observed flux. 
\cite{hen+98} imaged the region at 1.3~mm using the IRAM 30~m telescope.
They derived the existence of an extended
component of $\sim$53$"$$\times$49$"$ and a total flux of 0.25~Jy, and a compact component, $<$13$"$ with a flux of 0.11~Jy. The ISO 200~$\mu$m flux and the SCUBA 850~$\mu$m fluxes are also expected to be dominated by the emission from the extended component.
At wavelengths larger than 100~$\mu$m, we
only consider the interferometric observations for our fit.
Our millimetric data constrain the dust mass (0.004~M$_{\odot}$) and the dust maximum grain size to 1~$\mu$m in the midplane and 0.1~$\mu$m in the surface of the disk.
Assuming a standard gas/dust ratio of 100, the total dust+gas mass of the disk is 0.40~M$_{\odot}$. 
The disk size derived from the SED fitting is consistent with that estimated from the 1.3~mm PdBI images.

\subsection{LkH$\alpha$ 198}
LkH$\alpha$~198 (V* V633 Cas, IRAS 00087+5833) is a HAeBe
star located at 600 pc \citep{cha85}.
\cite{hil+92} classified 
LkH$\alpha$~198 as a A5 pre main-sequence star, while
\cite{her+04} found a spectral type of B9$\pm$2.5. 

A strong CO bipolar outflow is associated with this region,
but it is not centered on LkH$\alpha$~198, but to the northwest \citep{can+84,ballad83}. 
A small cluster of intermediate-mass young stellar objects is surrounding LkH$\alpha$~198.
V376~Cas is another HAeBe star located about 40$"$ north of LkH$\alpha$~198. 
An infrared companion (LkH$\alpha$~198--IR) was detected 
$\sim$6$"$ north of LkH$\alpha$ 198 by \cite{lag+93}. A submillimetre source
(LkH$\alpha$~198--mm) was identified 19$"$ northwest of LkH$\alpha$~198 
by \cite{sanwei94}. \cite{hajbas00} classified this source as an 
extreme Class I object. Extreme class I objects have been proposed by 
\cite{lad91} to refer to early class I objects that are seen through an
edge-on disk, thus explaining why they cannot be observed at NIR wavelengths.
BIMA observations by \cite{mat+07} revealed that the single-dish outflow
detected by \cite{can+84} is indeed the combination of the several bipolar outflows associated with
these stars. Evidences for the existence of bipolar outflows associated with V376~Cas and  LkH$\alpha$~198
were found.
One, even two outflows seemed to be associated with LkH$\alpha$~198, the most prominent in the CO emission 
with its axis in the northeast-southwest direction. A strong north-south outflow is associated with
LkH$\alpha$~198--mm. Although CO emission is detected toward LkH$\alpha$~198--IR, there is no evidence
of outflow in this IR source.

We did not detect millimetre emission toward LkH$\alpha$~198 at $\sim$1.3~mm and $\sim$3.0~mm 
Intense emission is found, however, associated with LkH$\alpha$~198--IR and LkH$\alpha$~198--mm (see Fig.~\ref{fig-maps} and Table~\ref{tab-lk}).
We used our 3$\sigma$ upper limit to the $\sim$1.3~mm emission to derive an upper limit to the mass of the disk associated with
this HAeBe star. This limit has been calculated assuming optical thin emission at 1.3~mm, that is the usual case in disks around
TT and Herbig Ae/Be stars, an average disk temperature of $\sim$40~K in agreement with the predictions of 
\cite{nat+00} for a B7 star, a dust opacity $\kappa$$_{1.3~mm}$=1.0 cm$^2$ g$^{-1}$ and a gas/dust ratio
of 100. We obtained that the gas+dust mass of the disk associated with LkH$\alpha$~198 is lower than 0.004~M$_\odot$.
 Alternatively we can have a small optically thick disk. In this case, our upper limit gives an upper limit to the
disk size but does not inform about its mass. Assuming that the disk average temperature
is $\sim$40~K, the radius of the disk should be $<$9~AU to be consistent with our 3-$\sigma$ upper limit to the 1.3~mm
emission.
LkH$\alpha$~198--IR and LkH$\alpha$~198--mm are two intermediate-mass young objects. LkH$\alpha$~198--mm
dominates the flux at 1~mm of the region and drives an energetic bipolar outflow. Therefore, the existence
of a circumstellar disk is expected. LkH$\alpha$~198--IR is supposed to be a more evolved object. 
The two sources have millimetre spectral indeces  $\sim$2.4, consistent with the emission arising from a disk and with the values measured for BD+61154 and RR~Tau.
In order to derive the disk masses,
we have assumed temperatures ranging from 15--40~K for the dust, that are reasonable values for disk associated 
with HAe stars.
We obtained that disk masses of 0.07--0.26~M$_\odot$, 0.42--1.42~M$_{\odot}$ for LkH$\alpha$~198--IR and
LkH$\alpha$~198--mm respectively.

\begin{table*}
\caption{Interferometric fluxes and masses for LkH$\alpha$198 field sources}             
\label{tab-lk}              
\begin{tabular}{l c c c c c c c}  \hline \hline \noalign{\smallskip}
Star                 &  RA(J2000)            &     Dec(J2000)            &    S(3.5~mm)      &  S(1.3~mm)  &  Spec. Index &\multicolumn{2}{c}{Masses (M$_\odot$)} \\ 
                 &    h:m:s          &   $^\circ$:$\arcmin$:$\arcsec$           &    mJy      &  mJy  &    &T$_d$=15~K  &  T$_d$=40~K  \\ \hline \noalign{\smallskip}  
LkH$\alpha$~198-mm   &  00:11:24.40     &   +58:49:43.71  &  10.9$\pm$0.1   &  132$\pm$5       & 2.5$\pm$0.1    &  1.42            &   0.42         \\
LkH$\alpha$~198-IR   &  00:11:25.87     &   +58:49:34.54  &  2.5$\pm$0.1   &  24.4$\pm$0.6  & 2.3$\pm$0.1   &  0.26            &   0.07       \\ \hline \noalign{\smallskip}
\end{tabular}              
\end{table*}

\subsection{Uncertainties in the disk mass, maximum grain size and disk outer radius}

 The assumptions made in the SED fitting process introduce some uncertainties in the values derived for the disk properties.

 The grain size distribution is usually taken to be $n(a) = n_0 \times a^{-n}$ for $a < a_{max}$, where $n_0$ is a normalization factor and $a_{max}$ is the maximum grain size. 
Experimental studies of fragmentation find that the value of  $n$ could vary between 1.9 for low-velocity collisions, and 4 for catastrophic impacts \citep{davrya90}.

\cite{tan+96} argued that the standard
$n$~=~3.5 power law is a very general result that depends only on the assumption that the fragmentation process is self-similar and that the collision rate varies as $a^2$. Simulations by \cite{tan+05}
of particle growth in stratified protoplanetary disks give vertically integrated size
distributions that at late times are well described by a power law $n$~=~3.0 for $a < a_{max}$, plus a population of much larger bodies. 
 The 1.3~mm/2.7~mm spectral index is considered a good indicator of the grain properties. In the Rayleigh-Jeans region of the spectrum, the dust emission is proportional to  $\nu^{(2+\beta)}$ where $\beta$ is the opacity index that can be accurately derived by fitting the observed SED. The value of $\beta$ in the submillimetre and millimetre ranges is an excellent indicator of the grain size distribution \citep{dra06}. 
There is some degeneracy between the index of the grain size distribution, $n$, and the value of the maximum grain size, $a_{max}$. An index of $n = 2.5$ means that a larger number of grains are closer to the maximum grain size than for a more typical value of 3.5. 
Assuming $n$~=~2.5 we would reach the same values of $\beta$ with grains slightly smaller. However, as previously discussed by several authors \citep[see e.g.][]{alo+08,ric+10tau}, 
for reasonable values of $n$, grain with sizes larger than few millimetre are needed to have values of $\beta$ lower than 1, as this is the case for RR~Tau.

 The dust composition in the midplane cannot be inferred from the SED. For this reason, we assumed the standard mixture. The silicate feature at $\sim$9.8~$\mu m$ provides, however, some information about the dust composition in the surface layer. In all of our sources, the silicate feature is present and can be fitted with the standard mixture. The absence of this feature, however, does not imply the absence of silicate grains but the lack of silicate grains at a temperature of $\sim$800~K.
The silicate grains could be either too far from the central source and thus not heated to temperatures sufficiently high to allow emission in the mid-IR, or these grains could be too large to be heated efficiently. The dust composition and maximum grain size in the midplane determine the value of dust emissivity at 1.3~mm, $\kappa_{1.3mm}$, and consequently has an influence on the derived dust mass. In Fig. 5 we show the value of $\kappa_{1.3mm}$ as a function of $a_{max}$ for different values of $n$ and different grain compositions. 
 For values of $a_{max}$ in the range 1~mm--~1~cm, all grain compositions and grain size distributions result in values of $\kappa_{1.3mm}$ that are  within a factor of 2--3 in agreement with the canonial value, 1~g$^{-1}$ cm$^{2}$.
For smaller grains,  $\kappa_{1.3mm}$  has a stronger dependance on the  grain composition and size distribution.
The variation reaches  at most a factor of 10.

 The outer radii calculated from the SED tend to be smaller than 
 those derived from the interferometric data fitting procedure.
 In the case of BD+61154 and VY Mon, the difference
 is within the modelling uncertainties and the results of our models are consistent with the observations. In the case of RR~Tau, 
 the outer radius calculated from the SED is definitely smaller. In the SED modelling we assume a surface density power-law truncated at the outer
 radius. Several authors have proposed a more realistic surface density distribution with a tapered exponential edge.
 This kind of surface density distribution gives smaller outer radii and therefore less consistent with our observations \citep{hug+08}.
 As discussed in Sect. 5.2, the large size measured by our interferometric observations in RR~Tau is likely due to a more complex
 geometry of the outer part of the disk, with the possible contribution 
 of a tiny envelope or some other extended structure such a secondary ring or spiral arms.
 Resolved images are required for a definitive answer.

\begin{figure}
\includegraphics{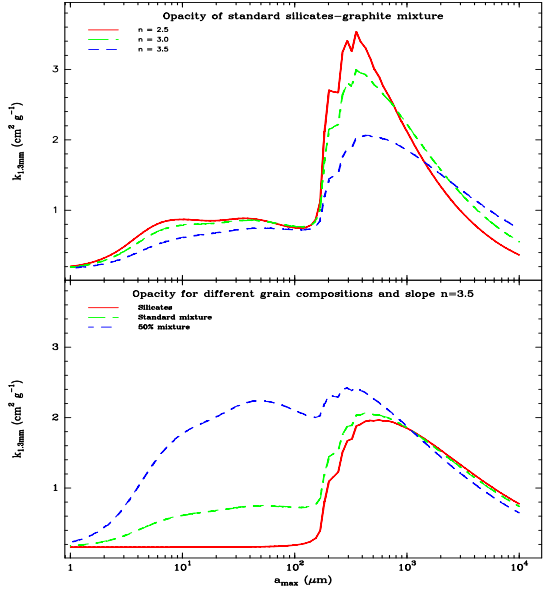}
\vspace{8.0cm}
      \caption{Dust opacity at 1.3~mm as a function of the maximum grain size, $a_{max}$, for different values of the grain size
distribution slope, $n$ ({\it top}) and for different dust compositions ({\it bottom}). Note that the variation of
$\kappa_{1.3mm}$ because of the different grain size distributions and grain composition is at most of 
a factor of $\sim$10.
}
         \label{fig-k1.3mm}
 \end{figure}

\section{Discussion}

We present interferometric data on 4 HAeBe stars with stellar masses in the 
range 3--7~M$_\odot$ and ages $\sim$1~Myr. All of them except LkH$\alpha$~198 have been detected and 
our SED modelling revealed that they are surrounded by 
massive disks (masses of about 0.03--0.4~M$_\odot$). Grains in these disks 
have grown only to sizes of $\sim$ $\mu$m--mm. Although grain growth is 
proceeding in these disks, 
their evolutionary stage is prior to the formation of planetesimals.
One possibility to explain the lack of disk detection in LkH$\alpha$~198 is 
that the age of this star is wrong because of the uncertain spectral type. 
\cite{tes+98} estimated
an age of 10~Myr assuming a spectral type of A4. If the star is older, most of 
the disk material could have been already dispersed by the star.
Another possibility is that the differences are due  to the vicinity of young protostars.
The UV radiation and bipolar outflows associated with these YSOs could have truncated the disk.

\begin{table*}
\caption{Interferometric observations in Herbig Ae/Be stars}             
\begin{tabular}{l c c c c c c c c c c c }  \hline
Star       &    M$_{star}$    &    Age   &  D &  $\lambda_1$   &  F$_{\lambda_1}$  & $\lambda_2$ &  
F$_{\lambda_2}$  & Spec. Index &  M$_{disk}$  &  Inter. & Ref  \\ 
           &   (M$_\odot$)    &  (Myr)   &  (pc) &  (mm)   &   (mJy)  & (mm) &  
(mJy) &    &   (M$_\odot$)    &   &   \\ \hline

Z~CMa       &  12      &    ...   &  930 & 2.7  &  3.0(0.7)  &  1.4  &  18(2)     &   2.7(0.9) &  0.12  & PdBI        & 1 \\
MWC~297     &   9      &    ...   &  250 & 2.6  &  79.6(5)   &  1.3  &   195 (2)  &   1.3(0.1) &   0.037  & PdBI, SMA & 1, 2 \\
R~Mon       &   8      &    ...   &  800 & 2.7  &  2.1(0.5)  &  1.4  &   8.7(2)  &    2.1(0.9)  &  0.009  & PdBI      & 1, 3, 4 \\       
VY~Mon      &   7      &  0.24 &  800 &  3.4  &  1.4(0.5)  &  1.4   &  46.1(1.4) &  3.9(1.5) &  0.30   & PdBI         & this work \\ 
MWC~614     &   5      &  0.5   &  240  &   2.6 &  7.6(1.4) &  1.3  &  70.8(7)   &  3.0(0.8) &   0.04  & OVRO         &  5 \\  
BD+61154   &  3.9      &  1.4  &  650 &  3.0  &  1.7(0.3)  &  1.1   &  33(1.0)   &  3.0(0.6) &  0.17   & PdBI         & this work \\ 
RR~Tau     &  3.6      &  1.3  &  800 & 3.5  &  2.1(0.2)   &  1.3   &  17.9(2)   &   2.2(0.4)  &  0.19   & PdBI       & this work \\
PV~Cep     &  3.5      &  0.5   &  500  &  2.7  &  32(1)     &  1.3   &  280(4)   &  3.0(0.1)   &   1.14   & CARMA    & 6 \\    
VV~Ser     &  2.7   &  3.9  &  260 & 2.6  &  0.7(0.2)  &  1.3   &  1.4(0.4) &   1.1(0.6)  &  0.0012  & PdBI           & 7 \\ %1.7x0.8 
AB~Aur    &  2.6   &  3.9   &  144  &  2.8  &  11(2)  &  1.4   &  85(5) &  2.9(0.7)   &   0.03 & PdBI                 & 8  \\    
HD~34282   &  2.5   &  5.6   &  400  &  3.4  &  5.0(0.3)  &  1.3   &  110(10)  &  3.2(0.5)   &   0.16  & PdBI         &  9, 10 \\ 
V892~Tau   &  2.5   &  5.0   &  140  &  2.7  &  39(0.5)   &  1.3   &  251(10)  &  2.5(0.1)   &   0.04  & CARMA        &  6 \\   
HD~163296  & 2.49  &  5.79  &  122 &   2.8 &  77.0(2.2) & 1.3  &  705(12)  &  2.9(0.2) &   0.13  & PdBI               & 11  \\ 
UX~Ori    &  2.4   &  4.2   &  450  &  2.6  &  3.8(0.4)  &  1.2   &  19.8(2)  &  2.1(0.4)   &   0.06  & PdBI          &  12  \\ 
WW~Vul    &  2.4   &  4.2   &  550  &  2.9  &  1.2(0.3)  &  1.2   &  9.1(1)   &  2.3(0.8)   &   0.04  & PdBI          &  12 \\  
MWC~480   &  2.2   &  5.98  &  130  &  2.8  &  35.2(0.8) &  1.4   &  235(4)   &  2.7(0.1)   &   0.12  &  PdBI         &  13, 14  \\    
MWC~758  &  2.17  &  6.26  &  200  &   2.6 &  6.7(1.3) &  1.3  &  56(1)  &  3.1(0.6) &   0.04  & PdBI                 &  15 \\       
CQ~Tau    &  1.5   &  5--10 &  100  &  3.4  &  13.1(0.5)   &  1.3   &  162(2)   &  2.6(0.2)   &   0.04  & PdBI        &  15, 16 \\ % 0.86x0.63
\hline
\end{tabular}              

\noindent
References: 
(1)~\citealt{alo+09};
(2)~\citealt{man+07}.
(3)~\citealt{fue+03}; 
(4)~\citealt{fue+06}; 
(5)~\citealt{mansar00}; 
(6)~\citealt{ham10}; 
(7)~\citealt{alo+08};
(8)~\citealt{pie+05};  
(9)~\citealt{pie+03}; 
(10)~\citealt{nat+04}; 
(11)~\citealt{ise+07}; 
(12)~\citealt{nat+01};
(13)~\citealt{pie+06}; 
(14)~\citealt{ham+06}; 
(15)~\citealt{cha+08};
(16)~\citealt{ban+11}; 
\end{table*}

\subsection{Correlation of the disk parameters with the stellar parameters.}

To investigate the possible dependence of the disk mass on the
stellar mass and age, we updated our previous compilation of disks (Table A.13 in Paper I). 
It is important to calculate the disk masses in a uniform
way. Similarly to Paper I, we calculated all disk masses, including our own sample, 
by assuming optically thin emission at millimetre wavelengths,
an average disk temperature that depends on the stellar spectral
type \citep{nat+00}, a dust opacity $\kappa$$_{1.3~mm}$=1.0 cm$^2$ g$^{−1}$, and a
dust spectral index $\beta$=1.0 in case we do not have $\sim$1.3~mm observations.
We subtracted the free-free contribution in the sources previously studied by our team 
\citep[this work;][]{fue+03, fue+06,alo+08,alo+09}

\begin{figure*}
\includegraphics{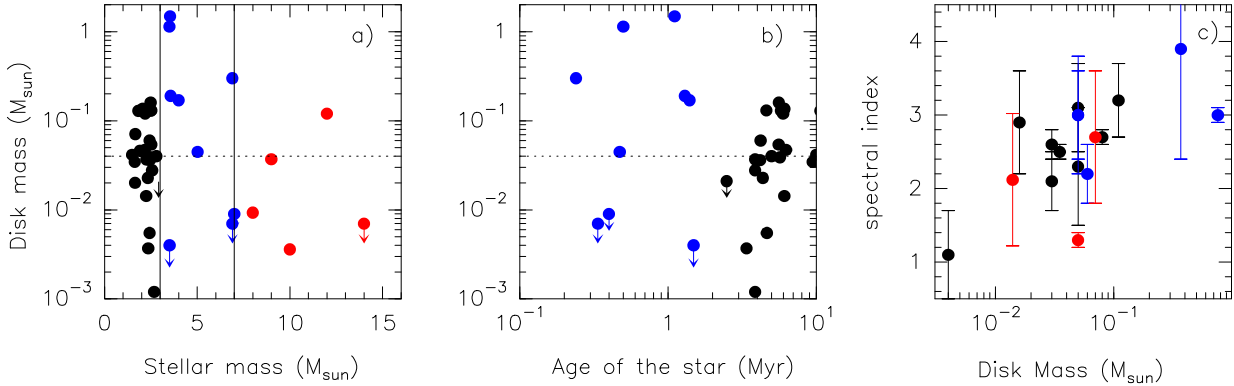}
\vspace{6.0cm}
      \caption{{\bf a)} Disk mass as a function of the stellar mass. The points are taken from the compilation of Paper I, updated with
the data presented in this work and the PV~Cep and V892 data published by \cite{ham10}.
Dashed line corresponds to $\sim$0.04~M$_\odot$,
the mean value of the disk mass for stars of  $<$3~M~$_\odot$. Solid contours indicates stars with masses of 3~M~$_\odot$ and 7~M~$_\odot$.
Different colours correspond to different ranges of stellar masses: from 1 to 3~M$_\odot$ (black), from 3 to 7~M$_\odot$ (blue)
and  $>$ 7~M$_\odot$ (red).
{\bf b)} Disk mass as a function of the age of the star for the same disks. {\bf c)} The $\sim$1.3~mm/$\sim$3.0~mm spectral index as a function of the
disk mass. Only disks with interferometric data at $\sim$1.3~mm and $\sim$3.0~mm are considered (Table 6).}
         \label{fig-disc}
 \end{figure*}

In Fig.~\ref{fig-disc}a, we represent the total disk masses
(gas+dust) as a function of the stellar mass assuming a gas/dust
ratio of 100. We only show in the plot the detections and
the meaningful upper limits (disk masses $<$0.04~M$_\odot$).
For stars with masses of $<$3~M$_\odot$, there is a cloud of points
around a mean value of $\sim$0.04~M$_\odot$. Stars with masses
of 3-7~M$_\odot$ usually have massive disks ($>$0.04~M$_\odot$).
But there are three stars without disk detections, and with upper limits
to the disk masses of $<$0.01~M$_\odot$.
In this range of stellar masses, most of the detections come from  
this work and correspond to young stars. This could bias our statistics to larger disk
masses. We need to complete our sample 
with older Ae stars to have a more realistic view of the disk evolution and determine the disk 
lifetimes in this stellar mass range.
Most massive stars ($>$7~M$_\odot$) have usually disk masses $<$0.04~M$_\odot$, consistent
with a more rapid disk dispersal as we concluded in Paper I.

It seems established that disk masses depend on the stellar mass because it determines
the photoevaporation efficiency and consequently, the disk lifetime. Disk masses should also depend on the
the stellar age because the star is expected to progressively disperse the gas and dust in the disk. 
In Fig.~\ref{fig-disc}b we plot the total disk masses as a function of the stellar ages.
Massive stars ($>$7~M$_\odot$) are not included in this plot because it is not possible
to have a reliable estimate of their age.
We do not find any correlation between the disk masses and the stellar ages. The first reason for this lack of correlation is that 
the star evolution strongly depends on the stellar mass. Massive stars evolves more rapidly than the lower mass ones.
We need to separate the stars in different ranges of stellar masses 
to have a meaningful plot. To investigate this effect, we have represented different ranges of stellar masses 
with different colors in Fig.~\ref{fig-disc}b. But even separating the stars in different ranges of stellar masses,
we do not find any obvious correlation. The first problem is that the number of stars in each stellar mass range is
very small, specially in the 3--7~M$_\odot$ range in which we only have 9 objects. Another problem is that the estimates of the
stellar ages are very uncertain. We have discussed in this paper the large uncertainties in the ages of
BD+61154 and LkH$\alpha$~198. The uncertainty in the ages of some stars is similar to the whole dynamic range of stellar ages.
Finally, the environment also influences the disk dispersal. Disks in clusters seem to be
less massive and more scarce than those around isolated stars \citep[see e.g.][]{manwil10}. The only way to investigate the influence of all these
factors in the evolution of the disks around HAe stars is to have more accurate estimates of the stellar ages and 
observe a large sample of disks that allow us to separate the sample
in different bins and make statistics in each individual bin.

\subsection{Correlation between grain growth and disk masses}

Since it is very difficult to find a correlation between the disk properties and the
stellar properties, we have tried to investigate the disk evolution by correlating two
disk parameters, in particular the 1.3~mm/2.7~mm spectral index and the disk mass.
The  1.3~mm/2.7~mm spectral index is a grain-growth tracer and can be understood
as an evolutionary indicator.
 \cite{ack+04} deduced from the FIR and submillimetre SED that the dust spectral index $\beta$  depends on the disk geometry (flared or flat disks). 
More recently, 
\cite{san+11} also found a correlation 
between the dust spectral index and the circumstellar mass measured on SCUBA maps. 
Although this result is encouraging, 
 it is based on single-dish observations, in which the disk and the enveloppe contributions are mixed,  and  from which the disk properties cannot be derived. 
We have made a compilation of interferometric observations of disks associated
to Herbig Ae/Be stars. 
Only the sources with  interferometric measurements at both $\sim$~3~mm and $\sim$~1.3~mm are
considered. 
In the case of CQ Tau, several measurements are available in the literature and  we used the fluxes measured by \cite{cha+08}.
More recent measurements by  \cite{ban+11}
resolve the disk and are not adequate for our statistical study of global properties.

In Fig.~\ref{fig-disc}c we show the millimetre spectral index as a function of the disk mass 
(calculated as explained in  Sect. 6.1). 
Disk with masses larger than 0.1 M$_\odot$  present a spectral index greater than 3, similar to those 
found in protostellar envelopes. 
However one should keep in mind that in these cases  the assumed dust opacity of 1 cm$^2$ g$^{-1}$ at 1.3~mm could be wrong, the real value being expected to be closer to $\sim$0.5 cm$^2$ g$^{-1}$. 
With this value, the estimated disk masses would be a factor of 2 larger.
Spectral indeces lower than 2 are associated with disks with low masses ($<$0.01 M$_\odot$).  
In these cases, centimetre-sized grains have grown in the midplane and their emissivity is smaller than the canonical value. 
The disk mass could thus be  underestimated as well. 
Disk masses in our sample range from 0.003 to 1.14 M$_\odot$, i.e. a factor of 380. 
Taking into account that at most a factor of 10 can be due to the unknown grain properties (see Sect 5.5), the range in our disk masses would be in anycase larger than 40. 
Under the reasonable assumption that the gradient in the disk mass is due to an evolutionary trend with the most evolved objects with the less disk mass, this means that only $<$2.5\% of the inital mass in dust grains will build up large grains. Another possibility is that  a significant fraction of the dust mass is locked in large grains ($>$~1~cm) that are not detectable in our observations.

\subsection{Comparison with TTs and Be stars}

 There are wealth of evidences of grain growth in disks associated with TTs. 
\cite{ric+10tau,ric+10rho} studied grain growth in  a large sample of disks in the Taurus and $\rho$ Ophiucus star forming regions. They found that the value of $\beta$ is quite constant with 
average value of 0.46. From their results, there is not correlation between the dust spectral index and the stellar parameters (mass, age). Althougth the stars in the Taurus region are older,
the stellar ages in the $\rho$ Ophiucus sample are similar to those of the stars in our Herbig Ae sample.
The characteristics found in the disks presented here are also very different from those observed in the disks around HBe ($\gtrsim$ 7~M$_\odot$) stars (see Paper I). 
Although the stars presented in this paper are older than the HBe stars presented in Paper I, their disks have larger masses, 
and the grains in the midplane have not reached centimetre sizes yet.

Dust coagulation models predict that grains grow to centimetre sizes in timescales of 10$^4$~yr \citep{zso+10}.
As particles grow, the velocity of the particles increases and the fragmentation becomes more efficient.
At some point, the steady state is reached and grain growth cease. The maximum size of the grains in the steady 
state depends on the gas and dust physical conditions (mainly the gas surface density), the turbulent velocity and the disk temperature 
(see \cite{bir+10,bir+11} for a comprehensive study).
The maximum grain size, $a_{max}$ increases with the gas surface density and turbulent velocity, but decreases with the disk temperature. Disks around HAe
stars are warmer than those around TTs. This could explain the larger grain sizes in TTs. 
Unless we assume that the turbulent velocity is larger in the disks around Be stars, the larger grain sizes in their circumstellar  disks is more difficult to explain.

Another possibility is that the steady state is not adequate for the studied stars. The stars in our sample are very young with a high accretion rate. \cite{man+06} showed that accretion in HAe stars decreases substantially at about 3~Myr. Accretion could change the grain size distribution resulting in an enrichment in small grains. Photo-evaporation could also affect the grain size distribution, specially in Be stars where the UV radiation is more intense. Gas and small grains are running away from the disk surface at radii larger than the gravitational radius. Radial drifts of large particles are not considered in these steady state models either.

\section{Summary and conclusions}
We completed our survey of disks around HAeBe stars by imaging the continuum emission at $\sim$3~mm and $\sim$1.3~mm 
toward BD+61154, RR~Tau, VY~Mon and LkH$\alpha$~198 using the Plateau de Bure Interferometer.
We have detected the disks associated with BD+61154, RR~Tau and VY~Mon with disk masses of 0.035~M$_\odot$, 0.05~M$_\odot$ and 0.40~M$_\odot$ respectively. 
The disk around LkH$\alpha$~198 has not been detected with an upper limit on the disk mass of 0.004~M$_\odot$. 
The modelling of the SEDs show that the grains in the mid-plane of the disks around BD+61154, RR~Tau and VY~Mon have sizes of $\sim$100, 1000 and 1~$\mu$m respectively. 
These massive disks are in the phase prior to the formation of protoplanets.
Their grains are smaller than those found around TTs and young HBe stars.

\begin{acknowledgements}
We are grateful to the IRAM staff in Grenoble (France) with their great help during the observations and data reduction.
This paper has been partially supported within the program CONSOLIDER INGENIO 2010, under grant  "Molecular  
Astrophysics: The Herschel and ALMA Era -- ASTROMOL" ( ref.: CSD2009-00038).
The research leading to these results has received funding from the European Community's Seventh Framework Programme (FP7/2007--2013) under grant agreement No. 229517 
\end{acknowledgements}

\bibliographystyle{aa} 
\bibliography{mnemonic,biblio-massdisk} 

\end{document}